\documentclass{aa}  
\usepackage{graphicx}
\usepackage{txfonts}
\newcommand{\kmos}{kms$^{-1}$}

\begin{document} 

   \title{Spectroscopic properties of 
\\ a two-dimensional time-dependent Cepheid model 
   \\ II. Determination of stellar parameters and abundances}

  \subtitle{}

   \author{V. Vasilyev\inst{1,2},
           H.-G. Ludwig\inst{1},
           B. Freytag\inst{3},
           B. Lemasle\inst{4},
            \and M. Marconi\inst{5}
          }
\institute{%
Zentrum f\"ur Astronomie der Universit\"at Heidelberg, Landessternwarte,
K\"onigstuhl 12, D-69117 Heidelberg, Germany \label{1}
\and
Max-Planck-Institut f\"ur Astronomie, K\"onigstuhl 17,
D-69117 Heidelberg, Germany, 
\email{vasilyev@mpia-hd.mpg.de}\label{2}
\and
Department of Physics and Astronomy at Uppsala University, 
Regementsv\"agen 1, Box 516, SE-75120 Uppsala, Sweden\label{3}
\and
Zentrum f\"ur Astronomie der Universit\"at Heidelberg, 
Astronomisches Rechen-Institut, Mönchhofstr. 12-14, D-69120 
Heidelberg, Germany\label{4}
\and
INAF -Osservatorio Astronomico di Capodimonte, Via Moiariello 16, I-80131
Napoli, Italy\label{5}
    }
   \date{Received ; accepted }
    \titlerunning{Spectroscopic properties of a 2D Cepheid model. II}
   \authorrunning{V.~Vasilyev et al.}  
  
\abstract{Standard spectroscopic analyses of variable stars are based
on hydrostatic 1D model atmospheres. This quasi-static approach
has not been theoretically validated.}
{We aim at investigating the validity of the quasi-static 
approximation for Cepheid variables. 
We focus on the spectroscopic   
determination of the effective temperature $T_\mathrm{eff}$, surface gravity $\log \,g$, 
microturbulent velocity $\xi_\mathrm{t}$, and a generic metal abundance $\log
\,A,$  here taken as iron.}
{We calculated a grid of 1D hydrostatic plane-parallel models covering the
  ranges in effective temperature and gravity that are encountered during the evolution
  of a 2D time-dependent envelope model of a Cepheid computed 
  with the radiation-hydrodynamics code CO5BOLD. We performed
  1D spectral syntheses for artificial iron lines
  in local thermodynamic equilibrium by varying the microturbulent velocity and
  abundance.  We fit the resulting equivalent widths to corresponding values
  obtained from our dynamical model for 150 instances in time, covering six
  pulsational cycles. In addition, we considered 99 instances during the
  initial non-pulsating stage of the temporal evolution of the 2D 
  model. 
  In the most
  general case, we treated $T_\mathrm{eff}$, $\log\,g$, $\xi_\mathrm{t}$, and
  $\log\,A$ as free parameters, and in two more limited cases, we
  fixed $T_\mathrm{eff}$ and $\log\,g$ by independent constraints.  We argue
  analytically that our approach of fitting equivalent widths is closely
  related to current standard procedures focusing on line-by-line abundances.}
{For the four-parametric case, the stellar parameters are
  typically underestimated and exhibit a bias in the iron abundance of
  $\approx-0.2\,\mbox{dex}$.  To avoid biases of this type, it is 
  favorable to restrict the spectroscopic analysis to photometric phases
  $\phi_\mathrm{ph}\approx0.3\ldots 0.65$ using additional information
  to fix the effective temperature and surface gravity.}
{Hydrostatic 1D model atmospheres can provide unbiased estimates
  of stellar parameters and abundances of Cepheid variables for particular
  phases of their pulsations. We identified convective inhomogeneities as the
  main driver behind potential biases. To obtain a complete view on
  the effects when determining stellar parameters with 1D models,
  multidimensional Cepheid atmosphere models are necessary for variables of longer
  period than investigated here.}

      \titlerunning{Spectroscopic properties of a 2D Cepheid model. II}
   \authorrunning{V.~Vasilyev et al.}  
  \maketitle

%
%________________________________________________________________

\section{Introduction}
Cepheids are some of the most  important variable stars 
in observational astronomy. 
First, the Cepheid period-luminosity (PL) relation 
\citep{1908AnHar..60...87L, 1912HarCi.173....1L}  is a powerful 
astrophysical tool to  measure distances within the 
Galaxy, and to measure cosmological scales \citep{2016ApJ...826...56R}.  
However, the chemical composition affects Cepheid pulsational properties, 
and it also reflects on the associated 
extragalactic distance scale,  even if no general 
consensus exists in the literature on the size and the sign of the effect. 
The accurate spectroscopic 
analysis of  \cite{2005A&A...429L..37R, 2008A&A...488..731R} 
shows that the metallicity affects the V-band 
Cepheid PL relation and that  metal-rich Cepheids appear to be 
systematically fainter than metal-poor ones 
at a fixed period, in agreement with theoretical prescriptions 
\citep{1999ApJ...512..711B, 2005ApJ...632..590M}.

Second, Cepheids are most convenient for a detailed study of radial
abundance gradients across the Galactic disk: abundances have been
derived for up to 25 elements from C to Gd \citep{2002A&A...381...32A,
2013A&A...558A..31L, 2015A&A...580A..17G, 2016A&A...586A.125D}.
Because they are bright supergiants, they allow us to probe the inner
disk \citep{2013A&A...554A.132G, 2015MNRAS.449.4071M, 2016MNRAS.461.4256A}
as well as the outermost regions \citep{2004A&A...413..159A, 2008A&A...490..613L, 2011AJ....142...51L}.

%Secondly, Cepheids are most convenient  for detailed  studies of radial 
%metallicity  gradients within the Galactic disk, and for setting 
%strong constrains on chemo-dynamical models of the Galaxy. 
%\cite{2002A&A...381...32A} measured 
%gradients  of 25 chemical elements 
%(from carbon to gadolinium)  in the solar neighborhood between 
%galactocentric distances in the range 6-11  kpc.
%The measurement  was extended toward  
%the galactic center \citep{2002A&A...384..140A}  
%and outer regions of the Galaxy \citep{2002A&A...392..491A,
%2003A&A...401..939L,2004A&A...413..159A}   up to $\approx 11$ kpc.  
%The radial iron abundance gradient was improved  by 
%\cite{2008A&A...490..613L}. They also showed an indication that 
%in the outer disk (10–12 kpc), the spread in metallicity depends on the
%Galactocentric longitude.  

Standard abundance determinations for Cepheids and non-pulsating stars
are based on grids of 1D hydrostatic stellar atmospheres. Calculations of
multidimensional Cepheid models are a sizable computational problem because of the different spatial and temporal scales.

During pulsations, the thermal structure of a Cepheid atmosphere changes.  The
effective temperature, gravity, and microturbulent velocity depend on the pulsation phase \citep{2002A&A...381...32A,2002A&A...392..491A,
2003A&A...401..939L,2004A&A...413..159A}, but the
metallicity does not (nor do individual abundances).  Standard grids of 1D
hydrostatic models cover the necessary wide range of stellar parameters
\citep{1992IAUS..149..225K, 2008A&A...486..951G}, typically adopting  
the mixing-length theory (MLT, see \citealt{1958ZA.....46..108B}) to describe convection.
\cite{1987VA.....30..197G} argued that for a Cepheid with a 10-day pulsational
period, deviations from hydrostatic conditions should be on a level of a few
percent. However, the thermal structure, which sets the conditions for the
line formation, differs from the corresponding thermal structure obtained in
hydrostatic models because of strong disturbances. Hence, the correctness of the
standard approach for determining stellar parameters and abundances
has to be validated.

In a previous paper (\citealt{2017arXiv170903905V}, hereafter Paper I), we have
introduced a 2D time-dependent Cepheid model calculated with the 
radiation-hydrodynamics code CO5BOLD \citep{2012JCoPh.231..919F}.

Here, we consider results of spectral syntheses taking the 2D
Cepheid model to provide artificial observational data, and apply the
standard approach to determine stellar parameters using 1D plane-parallel
hydrostatic model atmospheres.  Stellar parameters are exactly known for the
2D model. Thus, we can check the standard approach for biases on
parameters, including dependencies on the pulsational phase.  Since we only have
one multidimensional model at hand, our findings are restricted to the class
of short-periodic Cepheids.

The paper is organized as follows. Basic properties of the 2D 
Cepheid model are summarized in Sect.~2. The grid of 1D models is presented in
Sect.~3.  In Sect.~4 we describe the line list and compare it with line lists
that are applied in measuring Galactic metallicity
gradients. Features of spectral line profiles from the 2D model are discussed
in Sect.~5. Results for the determination of stellar parameters for the four-,
three- and two-parameter case are presented in Sect.~7.

\section{Two-dimensional Cepheid model}
Radiation-hydrodynamics simulations of a short-periodic 
2D Cepheid model in  Cartesian
geometry   employing gray 
radiative transfer  were calculated with the CO5BOLD code \citep{2012JCoPh.231..919F}.
The model has a nominal effective temperature of 
$T_\mathrm{eff}=5600$ K, a constant depth-independent 
gravitational acceleration  of
$\log\,g=2.0$, and solar metallicity.  
The model shows self-excited  pulsations presumably 
due to the $\kappa$-mechanism 
\citep{1917Obs....40..290E, 1963ARA&A...1..367Z}  
in the fundamental mode  with a pulsational period of $\approx2.8$ days.
The model features are a realistic treatment  of convection and radiative transfer.  
The construction of the model was a numerical challenge because of the 
extremely small time step imposed by the short radiative relaxation 
time in the employed time-explicit numerical scheme. It enforces the restriction to 2D models, 
at the moment.  A detailed description of
the physical and spectroscopic properties of the model, including the effects of the 
Cartesian geometry and  gray radiative transfer, can be found in Paper~I.

Starting with the initial hydrostatic condition, the 2D model reaches 
a quasi-stationary state with developed convection at $t \approx 4.5 \times 10^6$ s
 (see Fig. 2 in Paper I). Furthermore,  the model exhibits self-excited 
oscillations starting at $t \approx 6\times 10^6$ s. Ninety-nine 2D  snapshots between these instances in time were taken 
for the analysis of the hydrostatic non-pulsating regime.
For the analysis of the pulsating regime, radiation hydrodynamics simulations  
provided  150 2D snapshots  covering six full pulsation periods.

\section{Grid of 1D LHD models}

A grid of 1D plane-parallel hydrostatic models was calculated  using a
Lagrangian hydrodynamics code (here after LHD), 
which solves the set of 1D radiation-hydrodynamics equations in the
Lagrangian frame.  Convective fluxes and velocities were calculated according
to MLT.  In the present work, the mixing-length parameter $\alpha$ was fixed to
1.5 for all 1D models.  The actual value of the mixing-length parameter has a
minor effect on the photospheric temperature structure of the 1D models since
the the convective zone does not reach into the optically thin regions of the
giant stars considered here.

The radiative flux of 1D LHD models was calculated using gray opacities.
The radiative transfer equation was solved adopting the Feautrier
scheme \citep{1964SAOSR.167...80F}.  Opacities, the equation of state, and
the chemical composition were taken as for the 2D model.  The effective
temperatures of 1D models were varied between 4900\,K and 6000\,K in steps
of $\Delta T=100$\,K. They cover the effective temperature range that is
encountered during the temporal evolution of the 2D model.  The upper boundary
of  1D models of the grid was set to below $\log\tau_\mathrm{R}<-6$ to
contain the line-formation regions of the given line list, which is
described in the next sections.

The surface gravity $\log\,g$ of  1D LHD models was taken in the range 0.7 to
3.5  in steps of $\Delta\log\,g=0.2$ to cover the effective gravity
range -- including acceleration effects -- found in the 2D model. The
equivalent width (EW) of a spectral line depends on the physical conditions in the
line formation region of the stellar atmosphere.  The pressure is one of the
key quantities that influence the EW. This quantity, in turn, is
controlled by the surface gravity $\vec{g}$.  In stellar
atmospheres that are in  hydrostatic equilibrium, the pressure gradient, $\nabla p$, 
is balanced by the gravity force:
\begin{equation}
\frac{\vec{\nabla} p}{\rho}=\vec{g},
\end{equation}
where $\rho$ is the density.  In the 2D model, the dynamics adds to the purely
gravitational acceleration, which results in a total acceleration
that  has to be
balanced by the pressure gradient.  The effective
gravitational acceleration  is the sum of the effects of gravity and
kinematic acceleration, $\frac{d\vec{v}}{dt}$:
\begin{equation}
\frac{\vec{\nabla} p}{\rho}=\vec{g}-\frac{d\vec{v}}{dt}=\vec{g_\mathrm{\,eff}}.
\label{2ndlaw}
\end{equation}
In the 2D model, the 
gas and turbulent pressures  
$\vec{\nabla} p=\vec{\nabla} p_\mathrm{gas} + \vec{\nabla} p_\mathrm{t}$
contribute to the total pressure.
Since it is primarily caused by convection, the turbulent pressure $p_\mathrm{t}$ is
significant at the bottom of the photosphere. However,
the gas pressure gradient still provides the dominant contribution to the total pressure gradient even 
deep in the photosphere (see Fig. 5 in Paper I) and line-formation regions. 
Finally, the  effective gravity  
of the 2D model was  derived with the total pressure gradient.
We did not include turbulent pressure in the 1D models so that
$\nabla p=\nabla p_\mathrm{gas}$. Even if we had tried to do so in the
framework of MLT, it would have  had little effect since the convective zone is
restricted to layers below the photosphere. 

If we consider a Lagrangian mass shell following the mean vertical mass motion
in the line formation region, we find that it is subject to substantial
acceleration during the pulsations.  When the
direction of motion at the phase of maximum compression is reversed, it experiences an
acceleration of $\approx (1.0 \ldots 2.5)\,\vec{g}$.  In total, this
corresponds to an effective gravitational acceleration of
$\vec{g_\mathrm{\,eff}} \approx (2.0 \ldots 3.5)\,\vec{g}$, which in 1D models
would have to be balanced by the pressure gradient according to
Eq.~(\ref{2ndlaw}). For now, we recall that the line formation 
in a dynamical atmosphere occurs under a variable effective gravity.  While
kinematic effects have a rather obvious impact, spectroscopically determined
surface gravities can also be affected for other reasons. For instance,
according to \cite{1985A&AS...59..403S} and \cite{2003A&A...407..691K}, the
surface gravity of Procyon as deduced from the ionization balance is
inconsistent with an independent estimate obtained with astrometric methods.

For Cepheids, a mean gravity can be estimated without a spectroscopic analysis.
\cite{1965ApJ...142.1649G} theoretically  derived the "period-gravity" relation, 
\begin{equation}
P \sim g^{-1},
\end{equation}
where $P$ is the pulsational period.  The empirically 
calibrated  "period-gravity" relation of \cite{1988Ap&SS.150..357T}
 can be used to estimate the 
mean gravity,  and  this can be taken as a first
approximation to derive stellar parameters. According
to \cite{2002A&A...381...32A} (their Fig. 1), however, pulsations produce a scatter of 0.8 
around the mean value.

\begin{figure}
\includegraphics[width=\hsize]{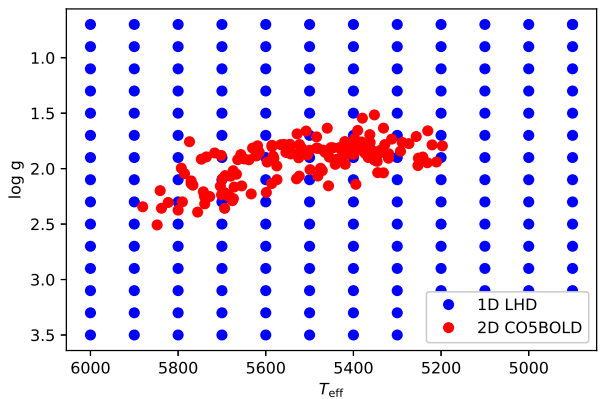}
\caption{Grid of 1D hydrostatic LHD models (blue circles) in the
  $T_\mathrm{eff}$--$\log\,g$ plane, and effective temperatures and
  gravities of the 2D model (red circles) for 150 instances in
  time, covering six full periods of pulsation.}
\label{grid}
\end{figure}

The full grid of 1D LHD models in the $T_\mathrm{eff}$--$\log\,g$ plane is
shown in Fig.~\ref{grid}. With different temperature and
gravity grid models, we intend to represent the states of the dynamical 2D atmosphere
encountered in the different pulsational phases.  It is clear that  
hydrostatic equilibrium is never exactly reached in the 2D model. The
timescale to attain  hydrostatic equilibrium is on the order of the
sound-crossing time, which is on the same order as the pulsational period.  We
might expect conditions with low effective gravity around the phase of 
maximum expansion.
At this point, the timescale of the convective instability
becomes longer than the pulsational period because of a low effective gravity,
which in turn decreases the efficiency with which the turbulence
is generated. 
The thermal structure is less perturbed by dynamical effects than in the 
phase of maximum compression.
\cite{2011ApJS..197...29F} found that photometric phases around $\approx 0.35$
(photometric phase zero corresponds to maximum light) are optimal for chemical
abundance analyses of RR~Lyrae stars with hydrostatic model
atmospheres. Recalling the differences between RR~Lyrae and Cepheid
variables (RR~Lyr stars have higher effective temperatures, higher velocity
amplitudes, shorter periods of pulsation, and exhibit stronger atmospheric
shocks), we expect similarities and address this point below.

\section{List of artificial iron lines}
For an accurate spectroscopic analysis, the line list and wavelength interval
have to be carefully selected.  Classically, the line list for the metallicity
determination contains isolated, unblended iron lines of different ionization
stages and excitation energies that cover a wide range of EWs.
The line strength depends on the physical conditions in the stellar
atmosphere. Thus, depending on the star, we have to include or exclude different
spectral lines in the analysis.

To keep our analysis general and to obtain a systematic overview, we did not
use a particular list of \ion{Fe}{i} and \ion{Fe}{ii} lines. Instead, our line list
consisted of 49 fictitious neutral and singly-ionized iron lines with a fixed
wavelength of $\lambda=5000$\,\r{A}.  Excitation energies of the \ion{Fe}{i}
and \ion{Fe}{ii} lines were taken to be $E_\mathrm{i}^{\ion{Fe}{i}}=1,3,5$\,eV
and $E_\mathrm{i}^{\ion{Fe}{ii}}=1,3,5,10$\,eV, respectively.  The oscillator
strengths were set to cover a wide range of EWs from 5\,m\r{A}
to 200\,m\r{A}.

\begin{figure}
\begin{center}
\includegraphics[width=200 pt]{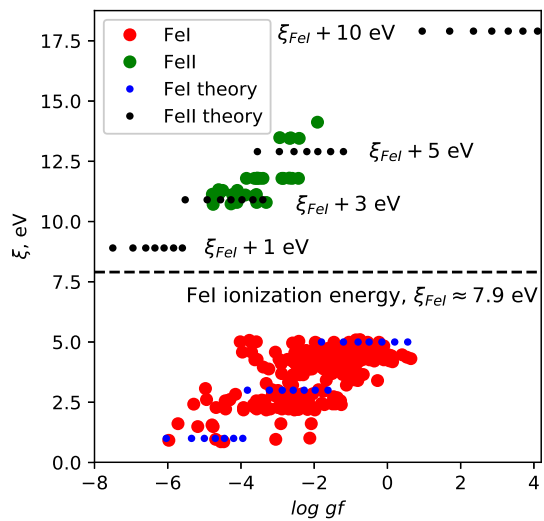}
\caption{Real (large circles) and our theoretical (small circles) 
   \ion{Fe}{i} and \ion{Fe}{ii}  lines in the excitation 
   energy--oscillator strength plane.\label{linelist}}
\end{center}
\end{figure}

Figure~\ref{linelist} shows the line parameters in the excitation 
potential -- oscillator strength plane.
We also plot a list of real
\ion{Fe}{i} and \ion{Fe}{ii} lines, which was used by
\cite{2007A&A...467..283L, 2008A&A...490..613L}, \cite{ 2010A&A...518A..11P},
and \cite{2013A&A...554A.132G,2014A&A...566A..37G}, to investigate the chemical
composition of Galactic Cepheids and measure the Cepheid metallicity gradient
in different parts of the Galactic disk.  
The group of \ion{Fe}{ii} 10\,eV lines is unrealistic 
with respect to lines used in observational analyses because of 
large  oscillator strengths and excitation potential.
Owing to the high-excitation energy, these lines form in deep regions of 
the atmosphere. As a consequence, they are strongly influenced by convection.
Especially during the maximum
compression phase, the convection is
amplified through the
high effective gravity (see Paper~I).  In order to be closer to the list of
real lines, and because of the strong sensitivity of \ion{the
Fe}{ii} 10\,eV lines to
dynamical effects, we did not consider these lines when we determined
the stellar parameters of the dynamical model. On the other hand, because
of the
infinite signal-to-noise ratio of our theoretical spectra, we did not exclude
the weakest lines with excitation potentials of 1, 3, and 5\,eV in the analysis.
The spectral synthesis was performed with
Linfor3D\footnote{http://www.aip.de/Members/msteffen/linfor3d/}
\citep{2017MmSAI..88...82G}
for the dynamical and 1D LHD models assuming local
thermodynamic equilibrium.

Before we describe our method for determining the stellar parameters and
present our results, we first discuss the features of the line profiles of the
dynamical model.

\section{Spectral line profiles of the dynamical 2D model  }
The homogeneous expansion or contraction of the atmosphere of a spherical
pulsating star transforms a symmetric Gaussian absorption line into a
characteristically asymmetric shifted line profile, as was shown
early on by
\citet{1919PNAS....5..417S}.  The asymmetry and shift depend on the radial
velocity of the line formation region \citep{2006A&A...453..309N}.  However,
the expansion or contraction of the atmosphere of a pulsating star is not
perfectly homogeneous.  This can, for instance, be observed in Balmer line
profiles of RR Lyrae stars \citep{2011AJ....141....6P} or is predicted by
global 3D radiation-hydrodynamics models of AGB stars
\citep{2017A&A...600A.137F}, and it is also demonstrated by our results in Paper I.
While in observations the line profile is observed averaged over the whole
stellar disk, the spectral synthesis for our 2D model provides information on
the variation in the spectral line profile in a spatially resolved fashion.

\begin{figure}
\includegraphics[width=\hsize]{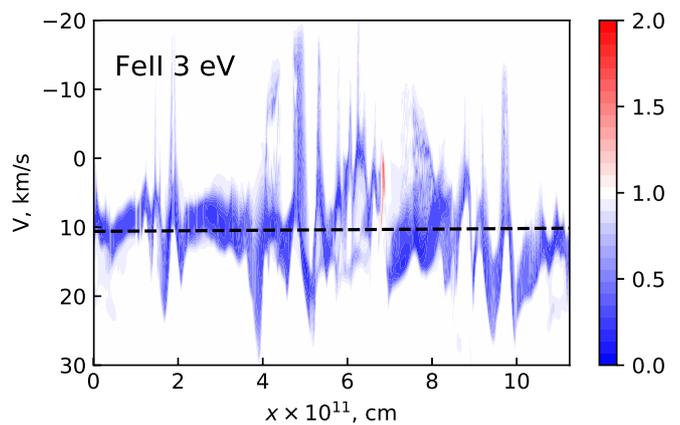}

\caption{Two-dimensional map of 
normalized line profiles of the strongest \ion{Fe}{ii} 3 eV line in
  terms of normalized intensity for
  a photometric phase of $\phi_\mathrm{ph}=0.56$. The horizontal 
  axis shows the horizontal spatial position within the modeled box. 
  The vertical axis represents
  the wavelength expressed as Doppler velocity.  Emission and absorption are
  indicated by red and blue lines,  respectively.  The continuum level is one,
  and it is shown in white.  The mean Doppler velocity is depicted by the
  dashed line.  The standard deviation of the Doppler shift velocities is
  6.1\,\kmos.}
\label{n=5}
\end{figure}

Figure~\ref{n=5}  shows the spatial variation along the horizontal position
in the modeled 2D box of
the normalized line profile of the strongest \ion{Fe}{ii} 3 eV line in terms
of intensity in vertical direction (inclination cosine $\mu=1$). The particular
instance in time corresponds to a photometric phase of 0.56, illustrating a
situation during the contracting phase of the 2D model. Absorption as well as
emission features are discernible. The line profiles vary
widely in terms of radial velocities and asymmetry, sometimes showing a
multi-component structure. The mean radial velocity is roughly 10\,\kmos\ for
the given example, with significant spatial variations caused by convective
inhomogeneities.

\begin{figure}
\includegraphics[width=\hsize]{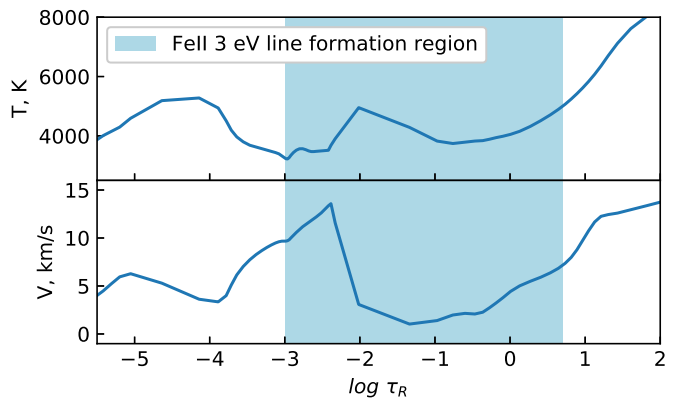}
\caption{Vertical thermal and radial velocity profiles of the section of the
  2D model in which the emission line profile of the strongest
  \ion{fe}{ii} 3\,eV line occurs.  The line formation region is indicated by the
  light blue area.}
\label{emission}
\end{figure}

Around a horizontal position  $x \approx 7.5\times 10^{11}$\,cm,  
emission lines are present.  Figure ~\ref{emission} 
shows vertical 1D thermal $T(\tau_\mathrm{R})$ and radial
velocity profiles around the location of emission.  We note that we use the
spectroscopic sign convention for the radial velocity, where a negative
velocity corresponds to a motion toward the observer and vice
versa.  The
\ion{Fe}{ii} 3\,eV line formation region is located at $\log \tau_\mathrm{R}
\approx -3 \ldots 0.7$.  Emission line profiles occur because of an inverse
gradient of the temperature at optical depths $\log\tau _\mathrm{R}
<-1.5$. The photospheric temperature of the thermal profile 
is $\approx 4000$\,K.  Local disturbances
of the thermal profile by convection can produce cold regions.  However, as
was shown in Paper~I, the variation in effective temperature of the mean 2D
model, which is the result of the horizontal averaging of the full 2D model at
fixed geometrical height, varies in a range between $5300$ and $5900$\,K.

The inverse temperature gradient and jump of the radial velocity at $\log
\tau_\mathrm{R}\approx -2$ in Fig.~\ref{emission} correspond to an accretion
front, where almost free-falling low-density material collides with a quasi-hydrostatically stratified deeper layer. The velocity difference between pre-
and post-shock regions is $\approx10$ \kmos,  which is substantial.  As was
described in Paper~I, we applied an artificial drag force at a number of grid
layers close to the top, reducing the velocities by a certain fraction per
time interval.  The drag force, which dampens waves, leads to a heating of the
top of the modeled box. This is visible at the temperature profile in
the regions where $\log \tau_\mathrm{R} < - 3.5$.  To minimize the effect of
the upper boundary, we considered lines that form at $\log
\tau_\mathrm{R}>-4$.

\section{Method of the parameter fitting}

One of the traditional ways to determine stellar parameters and elemental
abundances is based on minimizing the difference between observed and
synthesized line profiles by varying the model parameters of a precomputed
grid. However, in our analysis, we did not compare line profiles because, as
we showed in the previous section, the line profiles of the 2D
model are asymmetric, Doppler-shifted, and have a multi-component structure.
The line asymmetry depends on the pulsational phase, whereas 1D LHD line
profiles are perfectly symmetric and unshifted and thus can only poorly
represent the 2D profiles. For these reasons, we only matched the EW of the lines. In the 1D spectral synthesis, the microturbulent velocity
$\xi_\mathrm{t}$ was varied in the range 0.0\,\kmos\ to 7.0\,\kmos\ in steps of
0.5\,\kmos\ to cover a typical observational range \citep{2002A&A...384..140A}.
The same microturbulent velocity $\xi_\mathrm{t}$ was adopted for all lines
from the line list.  The result of the spectral synthesis is an array of EWs
$W_\mathrm{i}=f_\mathrm{i}(T_\mathrm{eff}, \log\,g, \xi_\mathrm{t})$, where
the index $i$ is a line number running from 1 to 42 for our set of artificial
iron lines.

Observationally, the metallicity of an observed star is an unknown parameter.
We mimicked this by including the iron abundance (metallicity), $\log\,A$, as
an additional parameter together with $T_\mathrm{eff}$, $\log\,g$, and
$\xi_\mathrm{t}$.  The EW of a spectral line depends on the product of the
abundance~$A$ and oscillator strength~$gf$, on a logarithmic scale $\log\,A
\ + \ \log\,gf$.  For a fixed $\log\,gf$, the abundance $\log\,A$ was changed
between $-0.6 \ldots 0.0$\,dex and $0.0 \ldots 0.6$\,dex in steps of
0.1\,dex.  Corresponding EWs were calculated using a cubic spline
interpolation in $\log\,W$-$\log\,gf$ space.

The EWs calculated with the grid of 1D LHD models were
finally used to best match the EWs from the spectral synthesis of the dynamical model.
We introduced a $\chi^2$ function to characterize the mean relative difference
between the EWs of the 2D and 1D models, which is in general a function of the
4D vector $\vec p=\big\{
T_\mathrm{eff},\ \log\,g,\ \xi_\mathrm{t},\ \log\,A \big\}$
\begin{equation}
    \chi^2( \vec p )= \frac{1}{N} \sum_{j} 
    \Bigg [\frac{W^\mathrm{2D}_\mathrm{j} - 
    W^\mathrm{1D}_\mathrm{j} (\vec p )}{f \cdot \sigma_\mathrm{Wj} } 
    \Bigg ]^2,
\end{equation}
where the sum goes over all iron lines of our line list, $N$ the total number
of lines, and $f$ is an arbitrary scaling factor for the assumed uncertainty
$\sigma_\mathrm{w,j}=W^\mathrm{2D}_j$. It should be understood that the
uncertainty stated above is not a statistical uncertainty, since our synthetic
lines are not subject to noise.  Taking the uncertainty as proportional to the
line strength itself was a convenient and reasonable way to express the
deviations in line strengths. We used a scaling factor $f=1,$ which in the
following allows a straightforward interpretation of a $\chi^2$ value: the
square-root of a given $\chi^2$ is the relative RMS deviation between 2D model 
and 1D model line strengths.

The best-fit combination of the parameters is found at the minimum of the
$\chi^2$ function.  We performed an exhaustive search of the minimum over our
grid of 1D models and took the model with minimum deviation as a first
approximation. To improve the location of the minimum, we interpolated between
grid points. We tested three different interpolation methods to further locate
the minimum in the 4D parameter space:
\begin{enumerate}
\item Radial basis functions (RBF) with a norm  
\begin{equation} \label{eq:rbfnorm}
\frac{1}{\sqrt{(\vec p /\Delta \vec p )^2 +1}}= 
\Big[ \sum_{i=1}^4 \Big ( \frac{p_\mathrm{i}}{\Delta p_\mathrm{i}} 
\Big)^2 +1 \Big]^{-1/2},
\end{equation}
where $\Delta \vec p = 
\big\{  \Delta T_\mathrm{eff},\Delta  \log\,g, 
\xi_\mathrm{t},  \Delta \log\,A  \big\}$  is the smoothing scale. 
The choice of the scale is based on the spacings in our grid of 1D models, which
are $\Delta T_\mathrm{eff}=100$\,K, 
$\Delta\log\,g=0.2$\,dex, $\xi_\mathrm{t}=0.5$\,\kmos, and 
$\Delta\log\,A=0.1$\,dex.
\item A fit of a quadratic form to reconstruct the 
shape of the function around the minimum.
\item A sequence of 1D cubic piecewise interpolations.
\end{enumerate}
Before we applied the methods for the analysis of EWs of the 2D model, they were
tested on EWs of one custom computed 1D model with
$T_\mathrm{eff}=5550$\,K, $\log\,g=2.0$\,dex, $\xi_\mathrm{t}=1.75$\,\kmos, and
$\log\,A=-0.05$\,dex, which is located between points of the grid.

\subsection{Test with radial basis functions}
The minimum point and its 80  nearest-neighbor points were taken  
 to describe the $\chi^2$ function with RBFs:
\begin{equation}
    \chi^2(\vec p)=\sum_{j=1}^{81} 
    \frac{\omega_j} { \sqrt{ \sum_{i=1}^4 
    \Big ( \frac{p_\mathrm{i} - p_\mathrm{ij}}{\Delta p_\mathrm{i}}
    \Big)^2 +1}},
\end{equation}
where the weights~$\omega_\mathrm{j}$ are results of 
the solution of the linear system of equations:  
\begin{equation}
    \sum_{j=1}^{81} p_\mathrm{ij} \omega_\mathrm{j} =  
    \chi^2(\vec p_\mathrm{i}), 
\end{equation}
with $p_\mathrm{ij}$ being the distance between the vectors
$\vec p_\mathrm{i}$ and $\vec p_\mathrm{j}$ in the parameter
space. % with the  norm given by Eq.~\ref{eq:rbfnorm}.
The reconstructed parameters for the test case are 
$T_\mathrm{eff}=5557$\,K, $\log\,g=2.03$ dex, 
$\xi_\mathrm{t}=1.77$ \kmos, and  $\log\,A=-0.035$ dex. 

\subsection{Test with a quadratic form}
The second interpolation method is based
on the reconstruction of the underlying $\chi^2$ 
function by a quadratic form. Again, 80 grid 
points from the hypercube around the  
minimum on the grid were considered 
to fit the  quadratic form%
\begin{equation}
f(\vec p)=\vec p^T\cdot A \cdot  \vec p +
B \cdot  \vec p +c ,
\end{equation}
where $A$ is a symmetric $4 \times 4$ matrix, and 
$B=\big\{b_1, b_2,b_3,b_4\big\}$ and $c$ are 
coefficients. The matrix $A$ has ten independent 
coefficients. The total amount of unknown parameters 
to fit is 15. After finding the  coefficients, multidimensional minimization 
algorithms can be used to determine the minimum of the function. 
The result of the best fit is 
$T_\mathrm{eff}=5551$\,K, $\log\,g=2.01$\,dex, 
$\xi_\mathrm{t}=1.72$\,\kmos, and $\log\, A=-0.002$\,dex 
 for the test case.

\subsection{Test with a sequence of cubic interpolations} 
The third method is based on using 1D cubic interpolations, splitting the
multidimensional interpolation in a sequence of 1D interpolations. The
4D interpolation demands 64 separate 1D
interpolation steps. The reconstructed parameters for the test case are
$T_\mathrm{eff}=5551$\,K, $\log \,g=2.00$\,dex,
$\xi_\mathrm{t}=1.75$\,\kmos, and $\log\,A=-0.045$\,dex.

Methods 2 and 3 are higher-order methods that work
better for regular data and when interpolation 
values outside the input range are needed.
The tests show that the data values 
are regular enough to make the higher-order
methods work better than the lower-order method 1.
The third method provided the closest match to the input parameters.  
Thus, we used the
cubic interpolation method.

\section{Results and discussion}
The parameters of the 2D model using the grid of 1D LHD
models were recovered for two different regimes of the temporal evolution of the 2D
model:
\begin{enumerate}
\item EWs for 99 instances in time were computed from  
the initial time evolution 
of the 2D model, 
when convection was the main dynamical 
driver and pulsations had not set in.
\item EWs for 150 instances were computed when pulsations 
had set in. They covered six full pulsational periods.
\end{enumerate}
The idea here is that by comparison of the two phases, we might be able to
separate effects of convection and pulsation.

\subsection{Four-parameter case}
This is our most general determination of the Cepheid parameters without using
additional constraints for fixing parameters. All information deriving from
the spectroscopic properties  is summarized here by the EWs of the
considered lines.  Each instance in time (``snapshot'') during the evolution
of the 2D model is considered independently.  This allows us to achieve an
understanding of systematic effects and the quality of the stellar
parameter determination as a function of the pulsation phase.
%\
\begin{figure*}
\centering
\includegraphics[width=17cm]{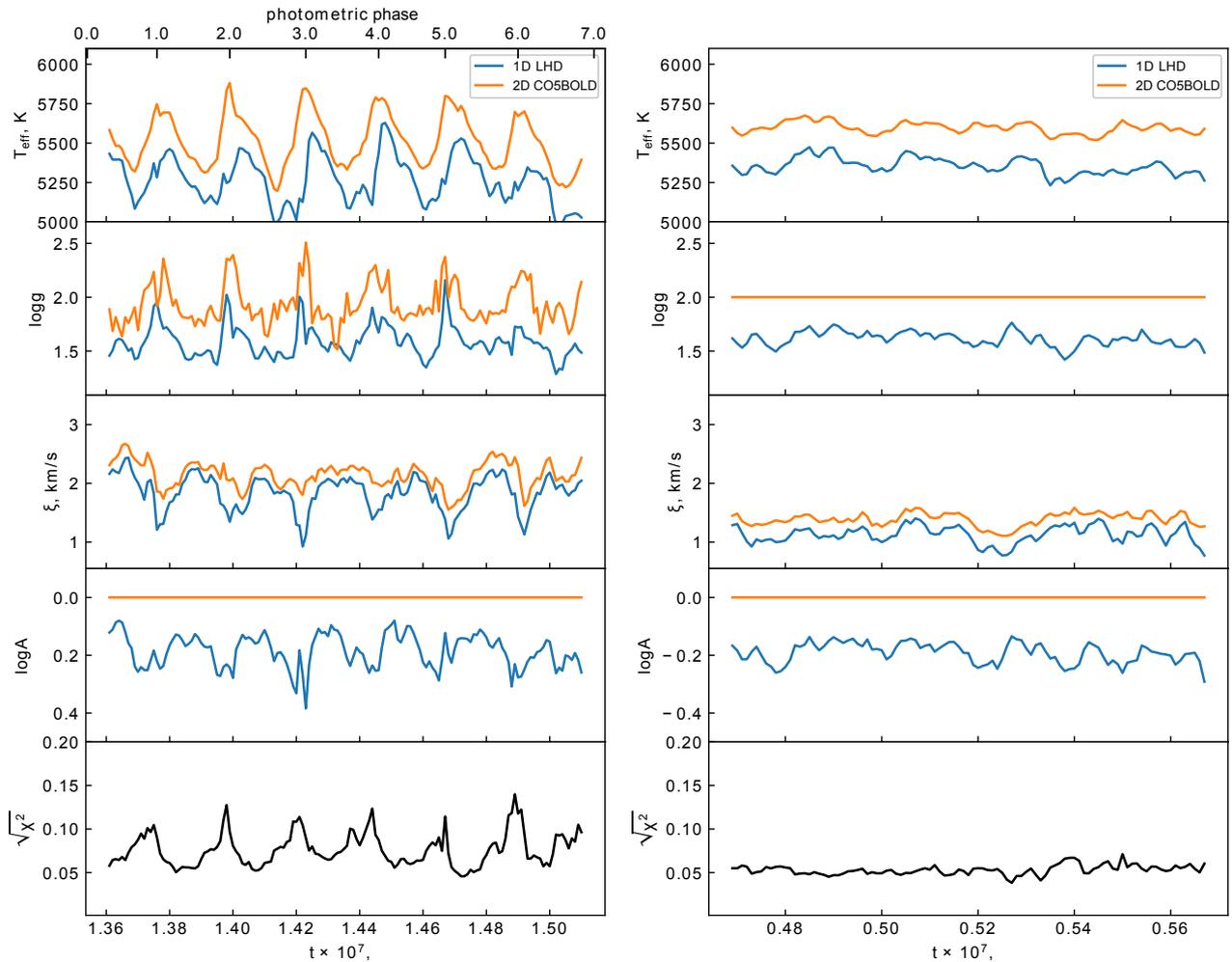}
\caption{Result of the four-parametric fitting for 
the pulsating (left panel) and non-pulsating (right panel) 
regimes. The parameters of the 2D model are shown by  orange 
lines for both regimes. The reconstructed parameters with the 1D LHD grid 
are shown by blue lines.  The relative RMS deviation $\sqrt{\xi^2}$ in  
line strengths  between the 2D  and 1D models is shown by the black line. 
}
\label{4param}
\end{figure*} 
%\
The results of the four-parameter fitting when pulsations have set in are shown in
the left panel of Fig.~\ref{4param}.  There are significant biases between the
2D model parameters and the result of the fitting using 1D hydrostatic
plane-parallel models.  The mean bias in the determination of the effective
temperature is $T_\mathrm{teff}^{2D} -T_\mathrm{teff}^{1D} \approx 250$\,K,
depending on the pulsation phase.  During phases of the maximum
compression, it is higher than 400-600\,K, whereas  it varies in a range from
100\,K to 200\,K for photometric phases $\phi_\mathrm{ph} \approx 0.3 \ldots 0.65$, which correspond to the maximum
expansion and early contraction stages.  Gravity and metallicity exhibit mean biases between 2D and
1D values of roughly $0.35$ and $0.2$, respectively.  The reconstructed
microturbulent velocity is slightly lower (by $\approx 0.1$\,\kmos) than the 2D
result, and shows a clear modulation with pulsational phase. We recall that the microturbulent velocity measured in the 2D case is not a
result of the standard spectroscopic measurement of $\xi_\mathrm{t}$ (see
details in Paper I), and neither is our fitting result based on least squares. 
In view of what to expect during an observational analysis, the bias
of the microturbulence should therefore be taken as only indicative.

As stated before, the square root of the $\chi^2$ function in
Fig.~\ref{4param} characterizes the mean relative difference between EWs of
the 2D and 1D LHD model.  Depending on the photometric phase, this quantity
varies from 5\,\% to 20\,\%. For the most extreme cases at maximum
compression, the difference is the largest.

It might be argued that pulsations mainly disturb the thermal structure of the
2D model, and physical conditions in the line-formation region
differ from the hydrostatic case.  To check this hypothesis, the fitting was
made for the case when pulsations have not set in.  The result of the fitting
is shown in the right panel of Fig.~\ref{4param}.  For this
regime, the relative difference in EWs is only 4-7\,\%, but the mean biases of the stellar parameters are very similar as in the time interval when
pulsations have set in. We conclude that the pulsations contribute additional
perturbations during the phase of maximum compression, but the main disturber
of the thermal structure is convection (see also Paper~I for further
discussion).

\begin{figure}
\includegraphics[width=\hsize]{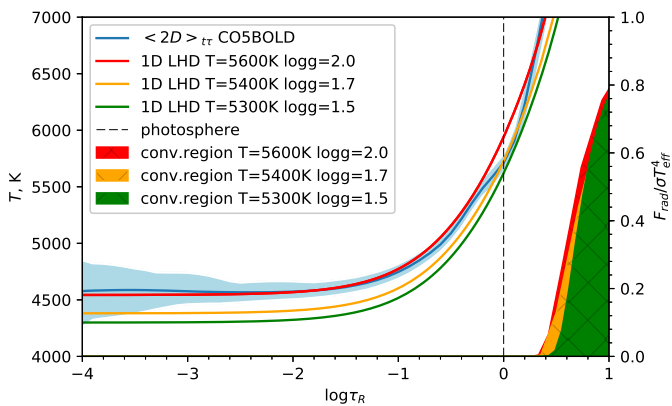}
\caption{Comparison of the thermal structures of the horizontally (on surfaces of
  constant optical depth) averaged 2D model 1D models for the case when
  pulsations have not set in.  The mean 2D thermal structure varies during 99
  instances in time within the light blue region. The temporally averaged 2D
  profile is shown in blue.}
\label{therm_struct}
\end{figure}

Biases in the determination of stellar parameters for the cases when
pulsations have and have not set in are caused by significant differences in
the EWs of the 2D model and 1D LHD models.  As we remarked above, this is
caused by the different thermal structures of these models in the
line-formation region.  Thermal structures of the mean 2D model, which are the
result of horizontally averaging over Rosseland optical depth, and 1D models
are shown in Fig.~\ref{therm_struct}.  The 1D models were chosen 
from the grid according to the results of the fitting for the non-pulsating regime,
taking  the effective temperature to be $T_\mathrm{eff}=5300,5400$\,K, and gravity 
 $\log\,g=1.5, 1.7$\,dex. Additionally, we  plot the 1D LHD model with 
 2D nominal surface gravity and effective temperature.  To simplify the discussion, 
 we only consider  the non-pulsating stage of the temporal evolution of the 2D model.
 
The temporal variation of the horizontally averaged structure of the 2D model
is indicated by the light blue region.  The photospheric temperature of the 2D
model changes as a result of convection. Convective overshoot and downflows
produce the bump in the thermal structure at $\log\tau_\mathrm{R}=0$.
Convective regions of the 1D LHD models are located below the photosphere, and
MLT cannot reproduce this particular thermal profile.  We have calculated
thermal profiles with different mixing length parameters, but there are no
qualitative improvements. The 1D LHD model with the same
effective temperature $T_\mathrm{eff}=5600$\,K and gravity $\log\,g=2.0$ as
the 2D model has appriximately the same temperature profile in a range of optical
depths from $\log\,\tau_\mathrm{R}=-3$ to $\log\,\tau_\mathrm{R}=-1$.
The 2D model has  a lower temperature for $\approx 250$\,K
 at $\log\,\tau _\mathrm{R}\approx 0$ than the 1D LHD model,
however.
The fitting
reconstructs the photospheric temperature.  Because it has the same 
photospheric temperature, the 1D model has  a cooler temperature
profile in the line formation regions  than  the 2D model.

For the EW $W$ of a weak line, the following expression
holds \citep{1992oasp.book.....G}:
\begin{equation} \label{eq:ew}
  W_\mathrm{\lambda} \sim 
  \int \frac{l(\lambda)}{\kappa(\lambda)} d\lambda,
  \end{equation}  
where $l$ and $\kappa$ are the line and continuous absorption coefficients,
respectively.  For the range of the effective temperature of the 2D model
between 5200\,K and 5800\,K, the negative hydrogen ion $\mathrm{H}^{-}$ is the
dominant source of continuous opacity, and it is very sensitive to
temperature.  Iron is mostly ionized in this temperature range.
Thus, neutral iron is a minority species, and its EWs 
increase with a decrease in effective temperature because the
$\mathrm{H}^{-}$ opacity drops (Eq.~\ref{eq:ew}). 
However, the behavior of  singly-ionized iron EWs as a function of 
temperature  is opposite  to that of neutral iron.  
For any snapshot of the non-pulsating sequence,
we can calculate EWs of the 1D LHD model 
taking the effective temperature, surface gravity, microturbulent 
velocity,  and metallicity of the  2D model. Our calculation  shows 
a mean relative difference  of 13\,\% between the 2D and 1D model EWs 
because the photospheric  regions of
$\log~\tau_\mathrm{R} \sim 0$ in the 2D model are cooler than in 
the 1D  LHD model.
Specifically, (i) the 1D LHD model EWs of \ion{Fe}{i} lines $W_\mathrm{\ion{Fe}{i}}^\mathrm{1D}$
are lower than $W_\mathrm{\ion{Fe}{i}}^\mathrm{2D}$,
$W_\mathrm{\ion{Fe}{i}}^\mathrm{1D}< W_\mathrm{\ion{Fe}{i}}^\mathrm{2D}$, and 
(ii) the EWs of \ion{the Fe}{ii} lines in the 1D model are larger than the 2D model EWs,
$W_\mathrm{\ion{Fe}{ii}}^\mathrm{1D} > W_\mathrm{\ion{Fe}{ii}}^\mathrm{2D}$.

From a physical point of view, EWs of \ion{Fe}{i} lines are most sensitive to changes of the effective
temperature, and rather insensitive to changes of the gravity. However,  EWs of \ion{Fe}{ii}
lines have the opposite behavior: they are sensitive to changes of the
gravity, and less sensitive to changes of the temperature. So, in a 
first qualitative fitting step, one has to decrease  
the effective temperature and surface gravity of the 1D LHD model 
to transform (i) and (ii) toward similar differences in 
relative EWs $(W^\mathrm{2D} -W^\mathrm{1D})/W^\mathrm{2D}$ for 
all ionization stages. The procedure increases
EWs of the 1D LHD model  with respect to $W^\mathrm{2D}$. As a result, 
one has to  decrease   the metallicity to reduce the difference $(W^{2D}-W^{1D})$
in a second step. 
It leads to a lower metallicity in the fit 
for the pulsating as well as non-pulsating phases.

\subsection{Three-parameter case}
In an observational analysis, the dimensionality  of 
the parameter space is typically reduced using additional information. 
\cite{1991PASP..103..439G} developed a method for  determining  the effective
temperature using the line depth ratios of pairs of weak lines of the same 
chemical element with two different excitation potentials for G and K dwarfs.
These ratios are  sensitive to the temperature variation  and  independent 
of metallicity effects for weak lines \citep{1994PASP..106.1248G}.  
\cite{2000A&A...358..587K} extended  this  method to derive precise 
temperatures of classical Cepheids and yellow supergiants with  
10-15\,K internal uncertainty using a calibration of the line depth
ratio versus effective temperature for 32 line pairs. 
\cite{2002A&A...381...32A, 2007A&A...467..283L,2011AJ....142...51L} used 
this calibration or the updated version \citep{2007MNRAS.378..617K} to derive 
the effective temperatures of Galactic Cepheids.

\begin{figure*}
\centering
\includegraphics[width=17cm]{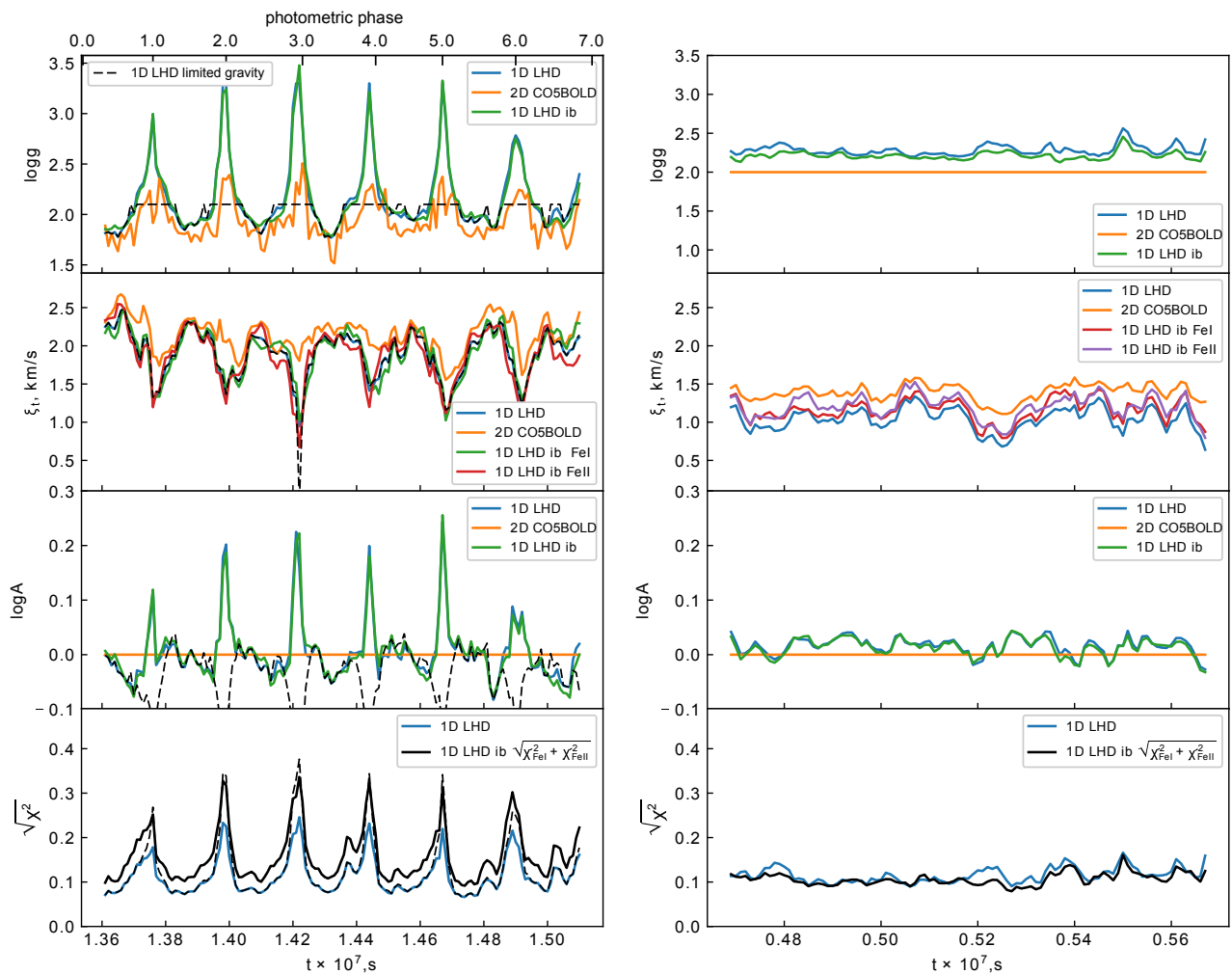}
\caption{Result of the three-parameter fitting for the pulsating (left
  panel) and non-pulsating (right panel) regimes. The parameters of 
  the 2D model are shown by  orange  lines for both regimes. 
  The reconstructed parameters with the direct three-parameter
  fitting using the 1D LHD grid  are shown by blue lines.
  Results using the ionization
  balance are shown by green, brown, and solid black lines.  For the pulsating
  regime, the result of the experiment with a gravity limit for the maximum
  compression phase is shown by the dashed black line.   The relative RMS deviation
  $\sqrt{\chi^2}$ in line strengths between the 2D  and 1D models is shown 
  by the solid blue and black lines for the direct three-parameter fitting and 
  for a fitting using the ionization balance condition, respectively. 
}
\label{3param}
\end{figure*}

The surface gravity can be estimated through the condition of the ionization balance, but for this, the correct effective temperature
is required because the
EW is quite sensitive to temperature changes.  Now we assume the
effective temperature to be known, and set to the value of the 2D model for
each particular instance in time. The effective temperature being fixed, we
estimate the three remaining parameters.  One-dimensional EWs were interpolated in the
$\log\,g$-$\xi_\mathrm{t}$-$\log\,A$ space for the fixed effective
temperature. The results of the three-parametric fitting for the non-pulsating and
pulsating stages of the temporal evolution are shown in the right and left
panels of Fig.~\ref{3param}, respectively.  The gravity estimate
in the
three-parameter fitting is based on the ionization balance, which is hidden
in the comparison of EWs of \ion{Fe}{i} and \ion{Fe}{ii}
lines. In the range of effective temperatures of the 2D model, most of the iron
is in the singly-ionized state.  Weak lines of \ion{Fe}{i} are insensitive to
pressure changes and, thus, to variations of the effective gravity.
Conversely, singly-ionized iron is pressure sensitive because
of the
opacity change in the negative hydrogen ion, which is sensitive to the electron
pressure, giving an overall dependence \citep{1992oasp.book.....G}
\begin{equation} \label{eq:grav}
W_{\ion{Fe}{ii}} \sim g^{-\frac{1}{3}}
.\end{equation}

In addition to the simultaneous fit in all three parameters, we also derived the
gravity by enforcing ionization balance.  For each instance in time, we
considered a two-parameter fit $\{\xi_t, \log \ A \}$ of the \ion{Fe}{i} and
\ion{Fe}{ii} lines separately, where the gravity was varied on the grid between
$0.7\ldots 3.5$\,dex.  The estimation of the surface gravity using  
the ionization balance condition was based on deriving the same 
abundance for \ion{Fe}{i} and \ion{Fe}{ii}.
Figure~\ref{3param} shows that the results of the fit using the 
ionization balance condition and the direct fit of all three parameters 
are in good agreement.

In comparison to the four-parameter case, the reconstructed microturbulent
velocity is very similar and consequently shows the same bias of
$\approx-0.1$\,\kmos.  The average microturbulent velocity estimated from the
ionization balance coincides with the result of the full three-parameter fit
for the pulsating and non-pulsating regimes.  The metallicities derived for
the non-pulsating regime closely correspond to the input value of the 2D
model. The convection effect leads to a small variation in $\log\,A$ with
time. It ranges between -0.03 and 0.04\,dex.

\begin{figure}
\includegraphics[width=9 cm]{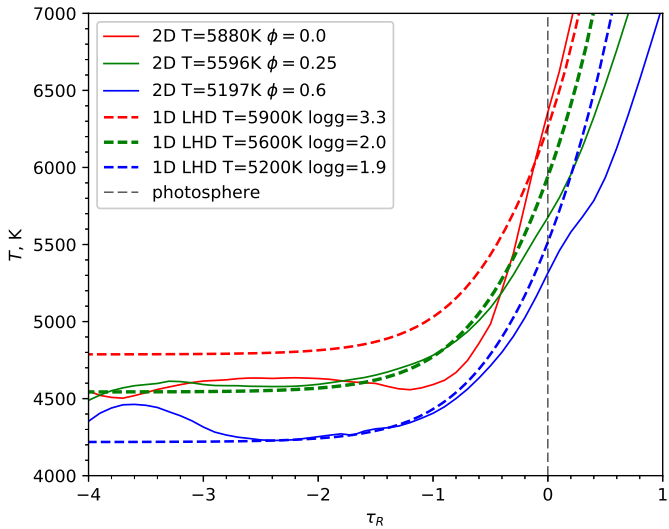}
\caption{Comparison of thermal structures of LHD models with the mean
  temperature profiles (average over optical depth surfaces) of the
  two-2D model for several photometric phases. $\tau_\mathrm{R}$ is
  the Rosseland optical depth.}
\label{therm_struct3}
\end{figure}

For the regime when pulsations have set in, there is a deviation 0.25\,dex in metallicity for the photometric phase with the maximum compression. Just
after the maximum compression phase, there is a time interval exhibiting the
smallest offset in metallicity stretching in the photometric phase between
$\phi_\mathrm{ph}\approx 0.3 \ldots 0.65$.  From a physical point of view, this
photometric phase coincides with the times of maximum expansion and subsequent
start of compression. During this period, the atmosphere is in a levitating
state, and convection and its disturbances of the thermal structure are not as
strong as during maximum compression.  The atmosphere is roughly in
hydrostatic equilibrium, and the 1D models reasonably reproduce the mean
thermal structure of the line formation region of the 2D model. This is shown in Fig.~\ref{therm_struct3} for the photometric phases 0.25 and 0.6.  It
leads to a correct reconstruction of the metallicity with a 10\,\% relative
difference between $W^\mathrm{1D}$ and $W^\mathrm{2D}$.

For the phase of maximum compression, the thermal structures of 1D and mean 2D
model in line formation regions differ appreciably despite the fact that they
have the same effective temperatures.  At optical depths $\log \,
\tau_\mathrm{R} \approx -1.3$, the mean 2D structure is 400\,K cooler than the
fitted 1D structure.  The 2D structure has a rather low resolution of the
optical depth scale around $\tau_\mathrm{R} \sim 1$ and exhibits steep
temperature gradients. For the 2D model, the horizontal averaging on surfaces
of constant optical depth yields lower temperatures for the photospheric
regions. However, experiments with increased resolution of the optical depth
scale using interpolation and subsequent horizontal averaging did not yield a
qualitative change of the mean 2D structure. This suggests that the
  horizontal inhomogeneities produced by convection and the horizontal
  averaging leads to the lower photospheric temperatures in comparison to 1D.
  In addition, the large changes in opacities on the coarse numerical grid
  and the steep temperature gradients might contribute as well. This has to be
  tested in future simulations with higher numerical resolution. 

As we described above, the \ion{Fe}{I} lines are insensitive to the surface
gravity. A high effective gravity of the 2D model leads to a decrease in
\ion{Fe}{II} EWs $W^\mathrm{2D}_{\ion{Fe}{ii}}$ according to
Eq.~(\ref{eq:grav}).  For the spectral synthesis with 1D LHD models of fixed
effective temperature, the surface gravitiy has to be increased
in a first qualitative fitting step to obtain the relative difference
$(W^\mathrm{2D}_{\ion{Fe}{ii}}-W^\mathrm{1D}_{\ion{Fe}{ii}})
/W^\mathrm{2D}_{\ion{Fe}{ii}}$ similar to \ion{Fe}{i}.  Owing to different
thermal structures of the line formation regions, 1D EWs are
smaller than the corresponding values of the 2D model.  As a second
step,  the iron abundance therefore has to be increased
to minimize the differences in
EW.  This leads to a positive bias for the metallicity during the
phase of maximum compression and to high values of the reconstructed gravity.

To understand the effect of the gravity on the metallicity bias during the phase
of maximum compression, we performed an additional test.  For this phase, the
gravity was limited to $\log\,g=2.1$ or less. The results of the fitting are
shown for the case when pulsations have set in in the left panel of
Fig.~\ref{3param}.  For a better fitting of the 2D EWs, the metallicity or microturbulent velocity can be changed.  However, EWs of weak lines
are insensitive to the change in microturbulent velocity. Thus, to reduce
the difference $(W^{2D}-W^{1D})$, the metallicity has
to be decreased.

\subsection{Two-parameter case}

We now assume that the effective temperature and the effective gravitational
acceleration are known from independent considerations.  We take the effective
temperature and gravity of the 2D model and interpolate EWs
calculated with the grid of 1D models in $\xi_\mathrm{t}$-$\log\,A$ space. In
this two-parameter case,  $\chi^2$ is a function of the microturbulent
velocity and metallicity alone.

\begin{figure*}
\includegraphics[width=\hsize]{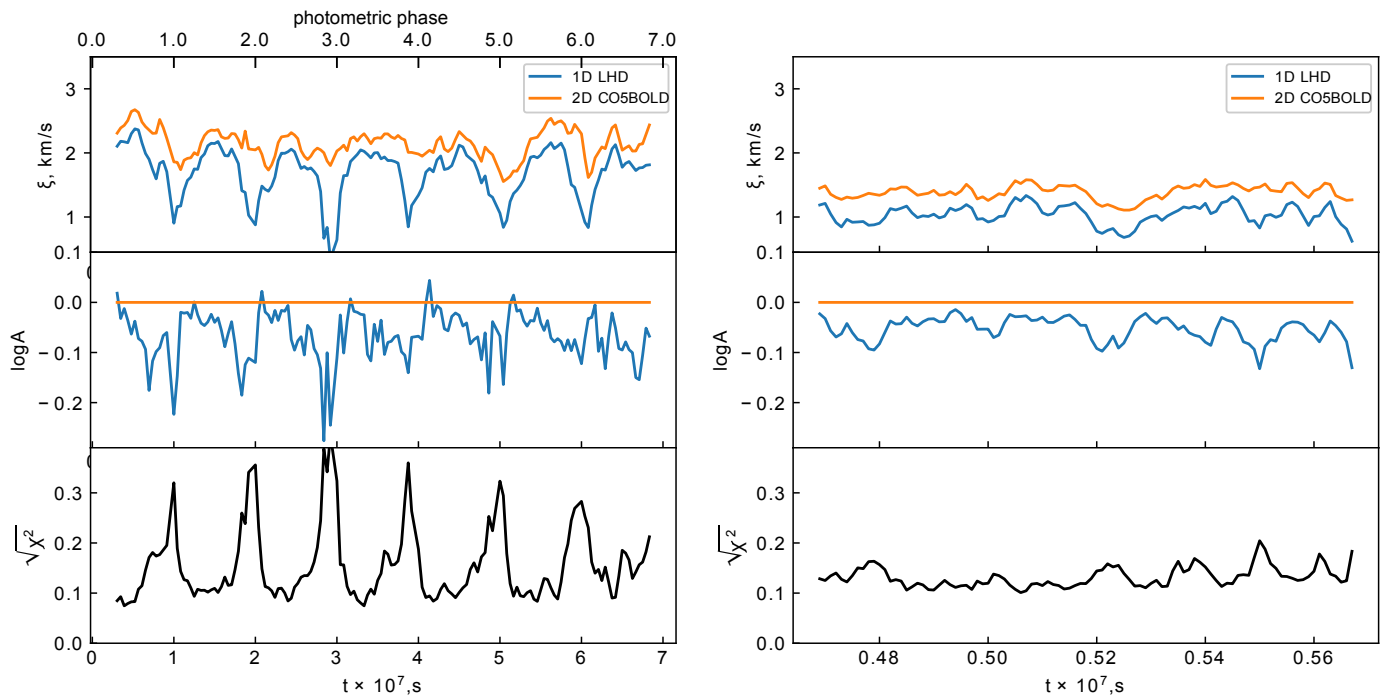}
\caption{The result of the three-parametric fitting for 
   the pulsating (left panel) and non-pulsating (right panel) 
   regimes. The parameters of the 2D model are shown by  orange 
lines for both regimes. The reconstructed parameters with the 1D LHD grid 
are shown by blue lines.  The relative RMS deviation $\sqrt{\chi^2}$ in  
line strengths  between the 2D  and 1D models is shown by black line. }
\label{2param}
\end{figure*} 
   
Again, we performed the two-parameter fit for the pulsating and non-pulsating
stages of the temporal evolution of the 2D model. The results are shown in the
left and right panels of Fig.~\ref{2param}.  The effective gravity of the 2D model for the pulsating case is
lower than the values of the three-parameter fitting.  As we
showed in the
previous subsection, the lower gravity leads to a negative bias of the
metallicity for most phases, except for the phase of the maximum expansion
and start of the contraction.  For this phase, the result of the fitting shows
a perfect reconstruction of the metallicity because the atmosphere is close
to hydrostatic conditions, and the 1D model can reproduce the thermal structure
(see Fig.~\ref{therm_struct}).  When pulsations have not set in,
the reconstructed mean metallicity is $\log\,A^\mathrm{1D}=-0.05$\,dex, which
is lower than the result of the three-parameter fit because the
gravity is fixed to $\log\,g=2$. According to Eq.~(\ref{eq:grav}), this leads on
average to higher EWs in 1D, and hence to a slight decrease of the metallicity.

Owing to the smaller number of free parameters in the two-parameter case, the
$\sqrt{\chi^2}$ value increases to a level of $\approx10$\,\%, except for the phases
of maximum compression.  When the analysis of the spectra is
performed by taking some
random photometric phase, a negative metallicity bias
of $\approx 0.06$\,dex is obtained on average.

\subsection{Connection between abundance and EW fitting}
We minimized a $\chi^2$ function of up to four parameters, 
which is the sum of relative differences of  the individual line 
EWs between the 2D model and 1D models. Here we wish to provide
arguments as to why the fitting in EW does not give a qualitatively
different result when the fitting of line-by-line abundances
is considered.
With fixed effective temperature and gravity, 
we reconstructed the thermal structure of the line formation regions.
The minimization of the mean difference of $W^\mathrm{2D}$ 
and $W^\mathrm{1D}$ EWs is equal to the minimization 
of the mean abundance differences $A^\mathrm{2D} -A^\mathrm{1D}$. 
This can be shown analytically for weak lines.  
For the two-parameter case, when $\chi^2$  is a 
function of the microturbulent velocity and abundance, we obtain 
\begin{equation} \label{eq:2par}
    \chi^2( A, \xi_\mathrm{t})=\frac{1}{N}\sum_{\mathrm {i}}^N 
    \left [ \frac{ W^\mathrm{2D}_\mathrm {i} 
    -W^\mathrm{1D}_\mathrm {i}( A, \xi_\mathrm{t})  }
    {f \cdot \sigma_\mathrm{W,i}} \right]^2,
\end{equation}
where N is the number of lines. 
To first order, the  difference in EWs 
of  the line $i$ between the 2D 
1D model can be represented by the Taylor expansion 
\begin{equation} \label{eq:taylor}
 W^\mathrm{2D}_\mathrm {i} - W^\mathrm{1D}_\mathrm {i} \approx
 \frac{\partial W_\mathrm {i}}{\partial \xi_\mathrm {t}} \cdot (\xi_\mathrm{t,i}^\mathrm{2D} 
 - \xi_\mathrm{t,i}^\mathrm{1D}) +
  \frac{\partial W_\mathrm {i}}{\partial A} \cdot (A_\mathrm{i}^\mathrm{2D} 
 - A_\mathrm{i}^\mathrm{1D}) +\ldots .
\end{equation}
The  uncertainty  $\sigma_\mathrm{W,i}$ in the EW 
is related with uncertainties in the abundance $\sigma_\mathrm{A,i}$ and  
microturbulent velocity $\sigma_\mathrm{\xi_\mathrm{t}}$ 
\begin{equation} \label{eq:error}
\sigma_\mathrm{W,i}=\sqrt{ \left[ 
\frac{\partial W} {\partial \xi_\mathrm{t}} 
                           \right ]^2 
\sigma^2_\mathrm{\xi_\mathrm{t}i} +\left[ 
\frac{\partial W} {\partial A} 
                           \right ]^2 
\sigma^2_\mathrm{A,i}}.
\end{equation}
Weak lines are insensitive to change in $\xi_\mathrm{t}$.  
Inserting the 
condition $\frac{\partial W} {\partial \xi_\mathrm{t}}=0$ 
into Eqs.~(\ref{eq:error}) and ~(\ref{eq:taylor}), we can modify 
Eq.~(\ref{eq:2par})
\begin{equation}
    \chi^2 \approx \frac{1}{N}\sum_{\mathrm {i}}^N 
    \left [ \frac{ A^\mathrm{2D}_\mathrm {i} 
    -A^\mathrm{1D}}{f \cdot \sigma_\mathrm{A,i}} \right]^2,
\end{equation}
which shows that the minimization of the difference of EWs
is equal to minimizing the difference in abundance between the 1D grid
and 2D model.

\subsection{Comparison of EWs}
\begin{figure}
\includegraphics[width=\hsize]{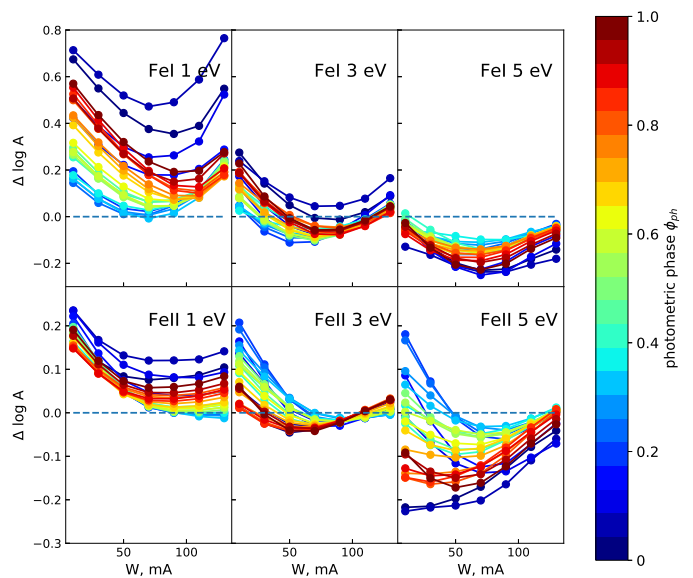}
\caption{Abundance corrections for individual lines for the best-fitting 1D models as a
  function of line strength and photometric phase.}
\label{grav_exper_ew}
\end{figure}

Using the curve of growths, we can transform the difference in EWs  $W^\mathrm{2D}-W^\mathrm{1D}$ into an abundance correction for each 
individual line. The individual abundance corrections 
as a function  of the line strength and photometric phase 
are shown in Fig.~\ref{grav_exper_ew} for the 
three-parameter case. 
As expected, the smallest abundance 
corrections correspond to photometric phases 
$\phi_\mathrm{ph}\approx 0.3 \ldots 0.65$. 
Generally, \ion{Fe}{i} lines have  larger abundance 
corrections than \ion{Fe}{ii} lines. 
Additionally,  within the same ionization stage, 
the abundance corrections become smaller 
with increasing excitation potential. 

\section{Summary and conclusion}
We have determined the stellar parameters (effective temperature,
gravity, microturbulent velocity, and metallicity),  of a 2D dynamical Cepheid model that was  calculated with the CO5BOLD code and has been presented 
in Paper~I.  
We took EWs based on the 2D model as observational data and
calculated a grid of 
the 1D plane-parallel hydrostatic model atmospheres for two regimes of the 
temporal evolution of the 2D model. We performed the
analysis  for three different cases: 
\begin{enumerate}
\item  A four-parameter case, where $\chi^2$ is a function of
  $T_\mathrm{eff}, \log\,g, \xi_\mathrm{t}$, and  $\log\,A$.  All reconstructed
  parameters are biased toward lower values than in the 2D model
  snapshots, and they are on average largely independent of the pulsations. The bias in
  the metallicity determination is $\approx -0.2$\,dex.

\item A three-parameter case, where the $T_\mathrm{eff}$ is fixed 
to the 2D value. $\chi^2$ is a function of
$\log\,g, \xi_\mathrm{t}$,  and  $\log\,A$.  Stellar parameters are determined with (i)
direct three-parameter fitting, and (ii) using the condition of ionization balance. 
The gravity estimate is higher than the effective 2D gravity for the
pulsating and non-pulsating regimes.  For the non-pulsating regime, the metallicity reconstruction agrees for all instances in time, whereas when
pulsations have set in, only the photometric phases  $\phi_\mathrm{ph}\approx 0.3
\ldots 0.65$ show a slightly biased metallicity estimate. 

\item A two-parameter case, where  $\chi^2$ is a function of
$\xi_\mathrm{t}$ and $\log\,A$. The metallicity estimate behaves
  qualitatively  similar to case~(2). 
\end{enumerate}

One-dimensional hydrostatic plane-parallel stellar model atmospheres employing different
MLT formulations generally cannot reproduce the mean thermal
structure of the 2D model for the whole range of optical depths. In
particular, the temperature at optical depth unity of the 1D models is typically higher than for
the mean 2D model. To avoid systematic biases in the determination of stellar
parameters of Cepheids with standard model atmospheres, we recommend analyzing
spectra taken during photometric phases $\phi_\mathrm{ph}\approx 0.3 \ldots
0.65$.

Our investigation was based on a single dynamical model of a Cepheid,
restricted to two spatial dimensions. A comprehensive theoretical
investigation of the line formation in the atmospheres of Cepheid variables would
require additional models,  in particular,
consideration of the full 3D case, and calculation of
2D models of higher numerical resolution  as well as  lower surface gravity, 
corresponding to longer pulsation periods.  
Including effects of
sphericity and investigating departures from local thermodynamic
equilibrium are desirable future improvements on the side of dynamical
multidimensional modeling. 
From a technical point of view, it would  be advantageous to use double precision 
because of the short radiative time step and, as a consequence of it, 
the  very small relative changes of the  model properties 
(particularly of  the internal energy) between time steps.

\begin{acknowledgements}
VV would like to thank Anish Amarsi for valuable comments on the draft.
HGL and BL acknowledge financial support by the Sonderforschungsbereich
SFB\,881 ``The Milky Way System'' (subprojects A4, A5) of the German Research
Foundation (DFG). The radiation-hydrodynamics simulations were performed at the
P{\^o}le Scientifique de Mod{\'e}lisation Num{\'e}rique (PSMN)
at the {\'E}cole Normale Sup{\'e}rieure (ENS) in Lyon.
\end{acknowledgements}

\bibliographystyle{aa} 
\bibliography{bibliography.bib}

\end{document}